\documentclass[twocolumn,
aps,nofootinbib,showpacs,showkeys,preprint
tightenlines,preprintnumbers,] {revtex4}

\usepackage{epsf,epsfig,subfigure,graphicx,amsmath,amssymb}
\usepackage{color}
\newcommand{\dis}[1]{\begin{equation}\begin{split}#1\end{split}\end{equation}}

\newcommand{\OVER}[1]{\,\overline{\hskip -0.5mm #1}}

\newcommand{\gev}{\,\textrm{GeV}}

\newcommand{\ie}{{\it i.e.}\ }

\newcommand{\Z}{{\bf Z}}
\newcommand{\CN}{{\cal N}}

\def\E6{{\rm E_6}}
\def\EE8{{\rm E_8\times E_8}}

\def\five{{\bf 5}}

\def\ten{{\bf 10}}
\def\one{{\bf 1}}

\begin{document}

\title{Abelian discrete symmetries $\Z_{N}$ and $\Z_{nR}$ from string orbifolds}

\author{Jihn E. Kim}
\affiliation{ Department of Physics, Kyung Hee University, Seoul 130-701, Korea
}

\begin{abstract}
We show that Abelian discrete symmetries $\Z_{N}$ and $\Z_{nR}$ from string orbifolds result by assigning
vacuum expectation values(VEVs) only to specified singlets. These singlets obtaining VEVs
carry the gauge charges (of the covering U(1) gauge group) as multiples of $N$ for $\Z_N$ and $n$ for $\Z_{nR}$.
We explicitly show this realization in a $\Z_{12-I}$ orbifold model.

\end{abstract}


\maketitle


\section{Introduction}

Discrete symmetries in string compactification are the key ingredients in
investigating effective low energy interactions \cite{DiscreteR, Kobayashi07}. It became more important after
realizing that gravity does not necessarily preserve global symmetries \cite{GravityGlobal, Kim13} such as the Peccei-Quinn(PQ) symmetry \cite{PQ77, KimRMP10}. At field theory level with supersymmetry(SUSY),
there exists a host of literature on discrete symmetries \cite{StroWitten85, DiscrGauge89, Ibanez92, BanksDine92, Preskill91}.

Discrete symmetries dictate the classical interactions. Given the interactions, quantum corrections
may or may not break the discrete symmetries. To cancel quantum mechanical anomalies,
in an effective theory at low energy with an integer normalization with the smallest integer charge magnitude set to 1, the needed
constraints have been obtained in Refs. \cite{Ibanez92, BanksDine92}. If the constraints cannot be satisfied, it was argued
that there exist new fractionally charged states and an enlarged symmetry group at high energy \cite{BanksDine92}. In
string compactification, we can obtain all the massless spectrum below the compactification scale $M_c$. The Kaluza-Klein(KK) states
have masses above $O(M_c)$, and we anticipate that no quantum mechanical inconsistency of light fields results from the KK states. Indeed in Ref. \cite{BanksDine92}, no inconsistency was found from string compactification with the string modular invariance conditions satisfied. The KK states may have consequently an enlarged symmetry compared to that of the corresponding massless spectrum \cite{BanksDine92}, but the massless spectra from string compactification already satisfy the constraints \cite{Preskill91} and an enlarged symmetry may not be mandatory.

\begin{figure}[!t]
\begin{center}
\includegraphics[width=0.45\linewidth]{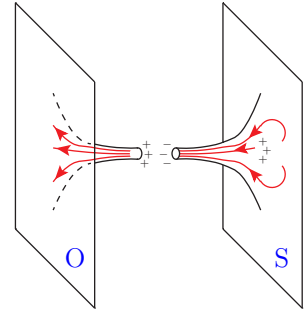}
\end{center}
\caption{Charges escaping through a wormhole. Cutting the flux lines through a narrow wormhole, the escaped gauge charges are recovered in the observable universe O. For the global charges, there is no flux lines and hence the escaped global charges are not recovered to O.} \label{fig:Wormhole}
\end{figure}

Since there does not exist a universally accepted quantum gravity at present,
here we comment only on the classical aspects of the gravity sector. Topology change
by black holes and wormholes can take out charges from our observable universe O, which may look breaking a symmetry or not from the point of O.  For example, as shown in  Fig. \ref{fig:Wormhole}, wormholes can take out charges from our universe O.
If the charge escaping O is the gauge charge then it drags the field lines which become a thin flux tube in the wormhole. If one tries to separate the shadow world S from O by cutting the wormhole, the vacant space between the cut surfaces must be null by creating + and -- gauge charges at the cut surfaces of Fig. \ref{fig:Wormhole}.  So, the observer O will notice that the gauge charge is conserved since he recovers the + charges of the cut surface, and the wormhole looks like carrying a hair. On the other hand, global charges have a different fate since they do not drag field lines. If global charges escape from the observable universe O to some shadow universe S, cutting the wormhole to separate the shadow world S from O does not leave any trace to O and the observer O notices that the global charge is not conserved in his universe O, concluding wormholes (or gravity) break global symmetries.

Around  a particle, let us consider topology for continuous and discrete symmetries. The continuous symmetry includes the gauge and global symmetries. Being continuous, the continuous symmetry is used to get all information around the particle through the Gell-Mann-Levy equation if the information at a nearby point is given, where the gauge and global continuous symmetries are distinguished by carrying flux lines or not. For discrete symmetries, string compactification allows discrete symmetries \cite{DiscreteR}. If discrete symmetries arise as discrete subgroups of $\EE8'$ of the hetrotic string, then they belong to the discrete gauge symmetry, which guarantees the unbroken symmetry by gravity \cite{DiscrGauge89}. Namely, we expect that the gravity sector does not break discrete symmetries.

Here, being not concerned with quantum gravity, we limit ourselves to the case of weak gravity. But for the gauge interactions, quantum corrections may break the discrete symmetry, which has been studied considering the discrete gauge anomaly \cite{StroWitten85, DiscrGauge89, Ibanez92, BanksDine92}. Considering the string compactification in this paper, however, we will not worry about
discrete anomalies if the discrete group arises from a subgroup of a gauge U(1) of string compactification. It is also known that the
anomalous U(1) conditions are always satisfied in string compactification \cite{Kim88}. In a sense, the field theoretic conditions presented in Refs. \cite{ChoiKW97, LeeRatz11} are automatically satisfied from field contents of string compactification.

The most discussed discrete symmetry is the R-parity in SUSY models \cite{Hall83}, allowing a CDM candidate.\footnote{The BNL $(g-2)^{\rm BNL}_\mu$ is difficult to explain in the MSSM with the exact R-parity. Note, however, that $(g-2)^{\rm BNL}_\mu$ can be explained with R-parity violation \cite{gminus2effSUSY} or in inverted effective SUSY with the R-parity \cite{IeffSUSY12}.
}
The R-parity is a discrete subgroup $\Z_{2R}$ of global U(1)$_R$. It is easily
understood in an SO(10) grand unification (GUT) with a spinor matter $\Psi={\bf 16}$ and a vector Higgs
$\Phi={\bf 10}$. The simplest lattice length of E$_8$ is $\sqrt{2}$,
\dis{
{\rm (Spinor~ lattice)}^2=2,~ {\rm (Vector~ lattice)}^2=2\,.
}
The spinor $\Psi$ in the group space of E$_8$ is represented by weights having half-integer entries,
\dis{
\Psi &= \left(\frac{\pm 1}{2}~ \frac{\pm 1}{2}~ \frac{\pm 1}{2}~ \frac{\pm 1}{2}~ \frac{\pm 1}{2}~ \frac{\pm 1}{2}~
\frac{\pm 1}{2}~ \frac{\pm 1}{2}\right),~ {\rm etc.},
}
and the vecor $\Phi$ is represented by weights having  integer entries,
\dis{
\Phi & = \left(0~ 0~ 0~ 0~ 0~0~ \pm1~ \pm1  \right),~ {\rm etc.}
}
Then, it is trivial to notice that the dimension-3 superpotential is $\Psi\Psi\Phi$ which allows
R-parity --1 for $\Psi$ and R-parity +1 for $\Phi$.  We can express this kind of $\Z_2$ parity by a
subgroup of U(1) whose generator is consisting of an odd number of $\pm 2$, for example,
\dis{
P=\left(0~ 0~ 0~ 0~ 0~\pm 2~ \pm 2~ \pm2  \right).
}
Thus, with $P\cdot \Psi={\rm odd}$ and $P\cdot \Phi={\rm even}$, we realize a kind of R-parity. The
basic reason for this comes from the weights having either half-integer entries or integer entries, \ie spinor or vector type representations in the group space of SO(10). In the low energy effective theory, the discrete group $\Z_2$ is unbroken if grand unification(GUT) scale vacuum expectation values(VEVs) are assigned only to those having even integers of $P$.
Without SU(5) singlet contribution, $P$ can be proportional to $3B-L$ for the fields in $\overline{\bf 5}$ of SU(5). But, adding an SU(5) singlet contribution $\propto B+L$, it can be made proportional to $B-L$ as noted in \cite{Lebedev08}.

\begin{widetext}

\begin{figure}[!t]
\begin{center}
\includegraphics[width=0.35\linewidth]{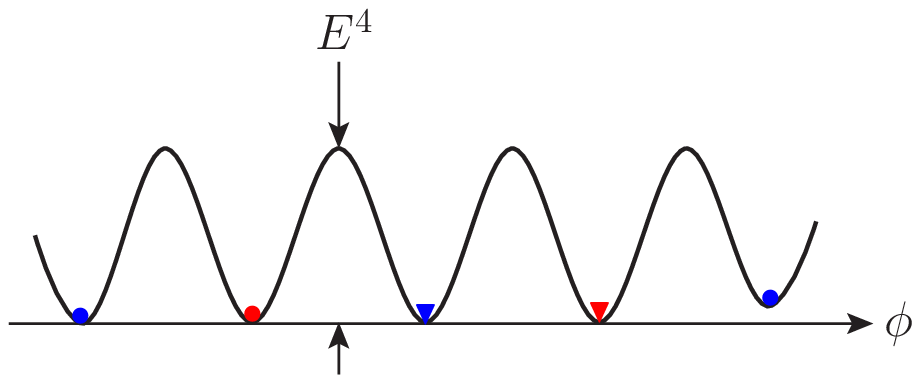}\hskip 1cm \includegraphics[width=0.35\linewidth]{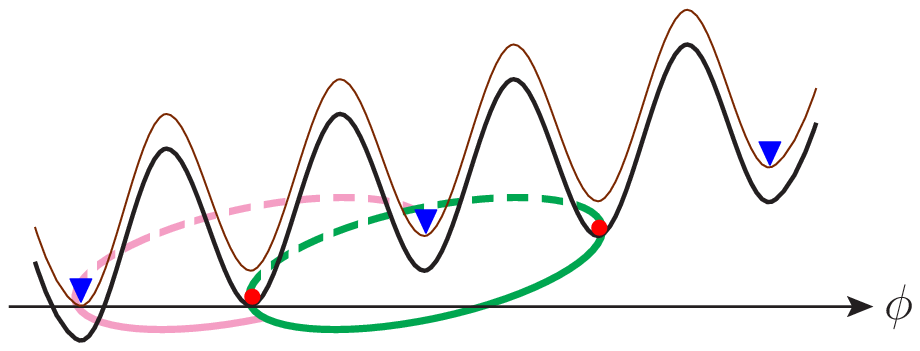}\\
(a) Gauge U(1)\hskip 5cm (b) Gauge and global U(1)'s
\end{center}
\caption{Spontaneous symmetry breaking of U(1). (a) The gauge symmetry broken to $\Z_4$. The height of $V$ ($E^4$) is expressible in terms of Higgs VEVs, and the vacua energies are identical. Higher order terms will break the shift symmetry. (b) For the gauge and global U(1)'s with an unbroken $\Z_2$. The height has the contribution from the anomaly and the global symmetry breaking higher order terms. The vacuum energy along the green and lavender loops (due to $\Z_2$) are tilted but $\Z_2$ is unbroken as the green
and lavender loops are identical except for their $\Z_2$ quantum numbers(marked as red bullets and blue triangles).} \label{fig:SSB}
\end{figure}

\end{widetext}

In contrast to the above $\Z_2$ subgroup of a gauged U(1), a discrete subgroup of global U(1)'s can be also considered: U(1)$_{\rm PQ}$ \cite{ChoiKimIW07, ChoiKS09} and U(1)$_R$ \cite{KapplR09}. In this case, the discrete subgroup can be approximate since the covering global U(1)'s are not exact anyway. Obtaining the scale of this kind of discrete symmetry breaking is model-dependent.  Of course, there is a possibility that some discrete subgroup of a gauged U(1) can be made exact.
For example, consider a spontaneously broken continuous (gauge or global) symmetry U(1). The vacuum energy in terms of
the variable $\phi$ developing VEV is shown in Fig. \ref{fig:SSB}.

For a supersymmetric CDM particle, an exact $\Z_{2R}$ is needed such that the lightest odd $\Z_{2R}$ can work as the supersymmetric CDM \cite{KamionPRP, AxinoDM}.

In this paper, we present this idea within a specific string compactification model SU(5)$\times$SU(5)$'\times$U(1)$^8$ already published in \cite{HuhKimKyae09}. Reference \cite{HuhKimKyae09}  is simple enough to present the basic idea because there is a small number of U(1)'s and hence a relatively small number of singlet fields and a discussion can be presented succinctly.

\section{$\Z_{n+2}$ versus $\Z_{nR}$}\label{sec:ZnZnR}

In the bottom up approach, firstly we may be interested in a discrete group of the minimal
SUSY standard model(MSSM). The anomaly consideration of discrete group with gauge fields and
gravity require certain conditions such that a need for light singlets beyond the MSSM fields may
result \cite{LeeRatz11}. Next we can consider the breaking scale of $\Z_{nR}$.

But in the top down approach, we do not worry about the anomaly constraints since all the
light particle spectra are determined such that no anomaly is present if we include the
Green-Schwarz mechanism \cite{GreenSch84,Kim88}.

The U(1)$_R$ symmetry is not exact since superpartner masses are split anyway due to SUSY breaking.
Thus, $\Z_{nR}$ symmetry is broken. In string compactification, it was argued that $\Z_{nR}$ arises from a discrete
remnants of the Lorentz group of the compactified space \cite{PetersenB09}.

In the effective theory of the MSSM, the discrete symmetries are found from the superpotential $W$.
Suppose that $W$ consists of cubic terms only, including those leading to the quark and lepton
Yukawa couplings. One possible R-charge assignment is that every field carries $R=2/3$ of $\Z_{2R}$. Another way to define the charges is non-$R$ symmetry $\Z_2$: $+1$ for quark and leptons and $-2$ for Higgs bosons. This kind of ambiguity from dimension-3 superpotential can be extended to $\Z_{nR}$ and $\Z_{N}$ (with $ N=n+2 $). Here, we consider the possibility of a discrete subgroup of U(1)$_R$ symmetry, in which case it is denoted as $\Z_{nR}$. For $\Z_{nR}$, the superpotential carries two units of U(1)$_R$ charge modulo $n$. For non-R symmetry, the superpotential carries zero unit of U(1) charge modulo $N$. If dimension-3 superpotential terms carry $n+2$ units of some U(1) charge, one can consider $\Z_{nR}$ if SUSY transformation is included or just $\Z_{n+2}$ if SUSY transformation is not included. For example, the $\Z_3$ orbifold does not allow the $\mu$ term and starts from dimension-3 $W$ \cite{Munoz93}. For $\Z_{2R}$, the R-charge assignment is $R=2/3$ for every field. For $\Z_{4}$, the R-charge assignment is $R=1$ for quark and leptons and $R=2$ for the Higgs fields, and it is required that some $R=4$ singlets obtain a VEV.

String compactifications do not restrict to dimension-3 superpotential terms only.
The selection rules are applied to all terms in the superpotential \cite{ChoiKimBk,IMNQ88,KimKyae07,ChoiKimIW07,SSMkimjh07,KapplR09}. But the number of possible dimension-3
superpotential terms allowed by the selection rules are generically smaller
than the number of dimension-$\ell$ terms for $\ell>3$, as checked in \cite{ChoiKimIW07}.
Therefore, it is likely that the discrete symmetry obtained from dimension-3 terms has an accidental symmetry larger
than the one obtained from all terms in the superpotential.

Consider a gauged U(1) direction in terms of the Cartan subalgebra orthogonal to the SM gauge group. Let the corresponding U(1) generator be $\Gamma$, and the charges are normalized such that the smallest nonzero integer units are $\pm 1$. One such example was given above as $P$. Assuming that dimension-2 term is not allowed, consider dimension-3 superpotential terms. If we cannot find a dimension-3 $\Gamma=0$ superpotential term consistent with the low energy gauge symmetry, then the next higher value of $\Gamma$ to consider is  dimension-3 $\Gamma=(n+2)$ superpotential terms.
If there exist such terms, then the discrete symmetry of dimension-3 superpotential is $\Z_{n+2}$. Restricting to dimension-3 superpotential terms, we can also declare that the discrete symmetry is $\Z_{nR}$. Now considering higher order terms, one finds that $\Z_{n+2}$ and/or $\Z_{nR}$ are broken.

The VEVs of singlet fields break some gauge symmetries down to $\Z_N\,(N=n+2)$ and $\Z_{nR}$.
If a singlet field carrying $(n+2)$ units of U(1) charge, the gauge group U(1) is broken to $\Z_{n+2}$.
If another singlet field carrying  $m$ units of U(1) charge, the unbroken discrete group is the largest common divisor of $n+2$ and $m$. This argument is extended to the VEVs of all singlet fields.
This simple fact is well-known for a long time, and recently it has been discussed systematically
in \cite{PetersenB09}. So, to find out the discrete groups, firstly we consider GUT scale VEVs of singlet fields. For the exact discrete symmetry, we must consider the electroweak scale VEVs of the Higgs doublets, $\{H_u, H_d\}$,  also. Next we consider the superpotential terms.

For the U(1)$_R$ symmetry, it is broken by the SUSY breaking effect, probably at the intermediate scale. But, the U(1)$_R$ breaking by the GUT scale VEVs is more conspicuous than the SUSY breaking effect below the intermediate scale \cite{Dine09,NelsonSeib94}.

In SUSY models, the $\mu$ term must be forbidden at the GUT scale \cite{KimNilles84}. A discrete subgroup of U(1)$_{R}$ must be approximate \cite{KapplR09, KapplR10, {SMZ2Z2}}. The R-parity must be a subgroup of a U(1)$_R$ group. With R-parity, however, the $\mu$ term is not forbidden. To forbid the $\mu$ term, recently, $\Z_{4R}$ which is a discrete subgroup
of U(1)$_R$ has been considered, where the quark and lepton fields are assigned 1 and Higgs doublets 0 charges of
$\Z_{4R}$ quantum number \cite{LeeRatz11}. Thus, $q u^c H_u$ and $q d^c H_d$, carrying two units of U(1)$_R$, are allowed but the $\mu$ term $H_uH_d$ is not allowed. In addition, dimension-4 superpotential term $qqq\ell$ is not allowed. Another way to forbid a GUT scale $\mu$ is the permutation symmetry $S_2\times S_2$ where a massless pair of Higgs doublets always appear \cite{Kim13}.

\section{Discrete groups from orbifold compactification}\label{sec:DiscGroups}

In the $\Z_{12-I}$ orbifold, the twist vector of complex 3-dimensions is
\cite{KimKyae07},
\dis{
\phi=(\frac{5}{12}~\frac{4}{12}~\frac{1}{12}),
}
which is associated with the boundary conditions of the right-moving superstrings, preserving only $\CN=1$ SUSY in 4D.
The gauge group space is twisted as well in terms of the shift vector $V$ and the Wilson line $W$,
\dis{
V &=\left(0~0~0~0~0~ \frac{-1}{6}~\frac{-1}{6}~\frac{-1}{6}\right)\, \left(0~0~0~0~0~ \frac{1}{4}~\frac{1}{4}~\frac{-2}{4}
\right)' \\[0.4em]
W &=\left(\frac{2}{3}~ \frac{2}{3}~ \frac{2}{3}~ \frac{2}{3}~ \frac{2}{3}~ 0~\frac{-2}{3}~ \frac{2}{3}\right)\,
\left(\frac{2}{3}~ \frac{2}{3}~ \frac{2}{3}~ \frac{2}{3}~ 0~\frac{-2}{3}~0~0 \right)'\,.\label{eq:DoubleSU5VW}
}
The gauge group and the resulting phenomenology of this model have been extensively discussed in \cite{HuhKimKyae09}.

\begin{widetext}

\begin{table}[t]
{\small
\begin{center}
\begin{tabular}{|c|c|c|c||c|c|c|c|}
\hline &&&&& &&\\[-1em]  $P+4V$ & $~\chi~$ & $(N^L)_j$ &
${\cal P}_4(f_0)$ & ~SU(5)$_X$~  & $B-L$& $ ~~{P}~~ $& $ ~~{Z_4, Z_8, Z_{10}, Z_{12}}~~ $ \\[0.5em]
\hline &&&&& &&\\[-1em]
$\left(\underline{+----}~;\frac{-1}{6}~\frac{-1}{6}~
\frac{-1}{6}\right)(0^8)'$ & L & $0$ &  $2$ &$2\cdot
{\bf 5}_{3}$ & $u^c:-\frac13 ,(\nu_e,e)^T: -1$  & $-1$& $-1$\\[0.5em]
$\left(\underline{+++--}~;\frac{-1}{6}~\frac{-1}{6}~
\frac{-1}{6}\right)(0^8)'$ & L & $0$ &  $2$ &$2\cdot \OVER{\bf 10}_{-1}$& $d^c:-\frac13 ,(u,d): \frac13, \nu^c:+1$& $-1$  &$-1$  \\[0.5em]
$\left(+++++~;\frac{-1}{6}~\frac{-1}{6}~\frac{-1}{6} \right)
(0^8)'$ & L & $0$ & $2$ &$2\cdot {\bf 1}_{-5}$& $e^+: +1$  &$-1$& $-1$ \\[0.5em]
\hline &&&&& &&\\[-1em]
$\left(\underline{1~0~0~0~0}~;\frac{1}{3}~\frac{1}{3}~
\frac{1}{3}\right)(0^8)'$ & L & $0$ & $2$ & $2\cdot {\bf 5}_{-2}$& $H_d: 0$ &$+2$ &$+2$   \\[0.5em]
$\left(\underline{-1~0~0~0~0}~;\frac{1}{3}~\frac{1}{3}~
\frac{1}{3}\right)(0^8)'$ & L & $0$ & $2$ & $2\cdot \overline{\bf 5}_{2}$ & $H_u: 0$ &$+2$ &$+2$ \\[0.5em]
\hline
\end{tabular}
\end{center}
\caption{The flipped-SU(5) chiral(L-handed) matter states in the $T_4^0$ sector  \cite{HuhKimKyae09}. The multiplicities of the fields in this table are 2. The color triplet and antitriplet of Higgs quintets carry $B-L=\pm 2/3$. The remaining family appears in the untwisted sector $U$. There, $\five_3(U), \OVER{\ten}_{-1}(U)$, and $\one_{-5}(U)$ carry $Z_i\,(i=4,8,10,12)= -1$ also.
}\label{tab:T4Ord}
 }
\end{table}

\end{widetext}

The U(1) generator we consider must commute with the unbroken gauge group generators of SU(5)$\times$U(1)$_X$. One such example is
\dis{
P=  (0~0~0~0~0~ 2~ 2~ 2)\, (0^8)'\,.
}
Since $Y$ and $X$ of the flipped SU(5) belong to the generators of SO(10), $B-L$  can be expressed in terms of $Y$ and $X$.
\dis{
B-L &=-4Y- X  \,.
}

Five U(1) gauge quantum numbers of two flipped SU(5) families appearing in Ref. \cite{HuhKimKyae09} are given in Table \ref{tab:T4Ord}. The $\Z_2$ subgroup of U(1)$_P$ can be a good candidate for the R-parity since the Higgs doublets are even and matter fields in Table \ref{tab:T4Ord} are odd under this $\Z_2$.  For an exact $\Z_2$, the GUT scale singlets having odd U(1)$_P$ quantum numbers should not develop VEVs.

U(1)$_R$ cannot be U(1)$_P$ since U(1)$_R$ is a global symmetry and U(1)$_P$ is a gauge symmetry.

\subsection{Singlets}

\begin{widetext}

\begin{table}[h]
{\small
\begin{center}
\begin{tabular}{|c|c|c|c|c||c|c|c|c|c|c|}
\hline &&&&&& &&&&\\[-1em]
 sectors & singlet states & $~\chi~$ & $(N^L)_j$ &
${\cal P}(f_0)$ & Label & $~~P~~$& $~~Z_4~~$& $~~Z_8~~$& $~~Z_{10}~~$& $~~Z_{12}~~$\\[0.5em]
\hline\hline &&&&&& &&&&\\[-1em]
$T_{4}^0$ & $\left(0~0~0~0~0~;\frac{-2}{3}~\frac{-2}{3}~\frac{-2}{3}
\right)(0^8)'$ & L & $0$ & $3$  & $S_1$ & $-4$ &  $-4$ &$-4$ &$-4$ &$-4$\\[0.5em]
$T_{4}^0$ & $\left(0~0~0~0~0~;\frac{-2}{3}~\frac{1}{3}~\frac{1}{3}
\right)(0^8)'$ & L & $1_{\bar 1},1_{2},1_{3}$ & $2,3,2$  & $S_2$ & 0& $-2$&$-2$ & $-2$ &$-2$ \\[0.5em]
$T_{4}^0$ & $\left(0~0~0~0~0~;\frac{1}{3}~\frac{-2}{3}~\frac{1}{3}
\right)(0^8)'$ & L & $1_{\bar 1},1_{2},1_{3}$ & $2,3,2$  & $S_3$ & 0&$0$&$0$ &$0$  &$0$\\[0.5em]
$T_{4}^0$ & $\left(0~0~0~0~0~;\frac{1}{3}~\frac{1}{3}~\frac{-2}{3}
\right)(0^8)'$ & L & $1_{\bar 1},1_{2},1_{3}$ & $2,3,2$ & $S_4$ & 0& $+2$&$+2$&$+2$  &$+2$\\[0.5em]
\hline &&&&& &&&&&\\[-1em]
$T_{6}$ &  $\left(0~0~0~0~0~;0~ 1~0\right) \left(0~0~0~0~0~\frac{1}{2}~\frac{-1}{2}~0 \right)'$ &
L & $0$ &  $2$ & $S_5$&$+2$&$+1$ &$-3$&$-5$ &$-7$ \\[0.5em]
$T_{6}$ &  $\left(0~0~0~0~0~;0~0 ~1\right) \left(0~0~0~0~0~\frac{-1}{2}~\frac{1}{2}~0 \right)'$ &
L & $0$ &  $2$  & $S_6$ &$+2$& $+3$&$+7$ &$+9$ &$+11$\\[0.5em]
$T_{6}$ &  $\left(0~0~0~0~0~;0~ -1~0\right) \left(0~0~0~0~0~\frac{-1}{2}~\frac{1}{2}~0 \right)'$ &
L & $0$ &  $2$ & $S_7$&$-2$& $-1$&$+3$ & $+5$ &$+7$\\[0.5em]
$T_{6}$ &  $\left(0~0~0~0~0~;0~0 -1\right) \left(0~0~0~0~0~\frac{1}{2}~\frac{-1}{2}~0 \right)'$ &
L & $0$ & $2$ & $S_8$&$-2$ & $-3$&$-7$ & $-9$ &$-11$\\[0.5em]
\hline &&&&& &&&&&\\[-1em]

$T_{3}$ & $\left(0~0~0~0~0~;\frac{-1}{2}~\frac{-1}{2}~\frac{-1}{2}
 \right)\left(0~0~0~0~0~\frac{3}{4} ~\frac{-1}{4}~ \frac{-1}{2}\right)'$ & L
 & $0$  & $1$ & $S_9$&$-3$&$-4$ &$-8$ & $-10$ &$-12$\\[0.5em]
$T_{3}$ &  $\left(0~0~0~0~0~;\frac{-1}{2}~\frac{1}{2}~\frac{1}{2}
 \right)\left(0~0~0~0~0~\frac{3}{4} ~\frac{-1}{4}~ \frac{-1}{2}\right)'$ &L& $0$  & $1$
 & $S_{10}$ &$+1$&$0$& $-4$ &$-6$ &$-8$\\[0.5em]
$T_{3}$ &  $\left(0~0~0~0~0~;\frac{1}{2}~\frac{1}{2}~\frac{-1}{2}
 \right)\left(0~0~0~0~0~\frac{-1}{4} ~\frac{3}{4}~ \frac{-1}{2}\right)'$ &L
 & $0$  & $1$ & $S_{11}$ &$+1$& $+2$& $+6$ &$+8$ &$+10$\\[0.5em]
$T_{3}$ &  $\left(0~0~0~0~0~;\frac{1}{2}~\frac{1}{2}~\frac{-1}{2}
 \right)\left(0~0~0~0~0~\frac{-1}{4} ~\frac{-1}{4}~ \frac{1}{2}\right)'$ &L
 & $1_{1},1_{3}$ &  $2$,$1$ & $S_{12}$&$+1$&  $+1$&$+1$& $+1$  &$+1$ \\[0.5em]
\hline  &&&&& &&&&&\\[-1em]
 $T_{9}$ &$\left(0~0~0~0~0~;\frac{1}{2}~\frac{1}{2}~\frac{1}{2}
 \right)\left(0~0~0~0~0~\frac{-3}{4} ~\frac{1}{4}~ \frac{1}{2}\right)'$ & L
 & $0$  & $1$ & $S_{13}$ &$+3$&$+4$& $+8$& $+10$ &$+12$ \\[0.5em]
$T_{9}$ &  $\left(0~0~0~0~0~;\frac{1}{2}~\frac{-1}{2}~\frac{-1}{2}
 \right)\left(0~0~0~0~0~\frac{-3}{4} ~\frac{1}{4}~ \frac{1}{2}\right)'$ & L
 & $0$ & $2$ & $S_{14}$&$-1$& $0$&$+4$ &$+6$ &$+8$ \\[0.5em]
$T_{9}$ &  $\left(0~0~0~0~0~;\frac{-1}{2}~\frac{-1}{2}~\frac{1}{2}
 \right)\left(0~0~0~0~0~\frac{1}{4} ~\frac{-3}{4}~ \frac{1}{2}\right)'$ & L
 & $0$  & $2$ & $S_{15}$&$-1$& $-2$&$-6$ & $-8$  &$-10$\\[0.5em]
$T_{9}$ &  $\left(0~0~0~0~0~;\frac{-1}{2}~\frac{-1}{2}~\frac{1}{2}
 \right)\left(0~0~0~0~0~\frac{1}{4} ~\frac{1}{4}~ \frac{-1}{2}\right)'$ & L
 & $1_{\bar 1}$,$1_{\bar 3}$ &  $1$,$1$ & $S_{16}$& $-1$&$-1$&$-1$  & $-1$ &$-1$ \\[0.5em]
\hline  &&&&&& &&&&\\[-1em]
 $T_{2}^0$ &$\left(0~0~0~0~0~;\frac{-1}{3}~\frac{-1}{3}~\frac{-1}{3}
\right)(0~0~0~0~0~\frac{-1}{2}~\frac{1}{2}~0)'$ & L & $2_{\bar1},2_{3}$ & $1,1$
 & $S_{17}$& $-3$&$-2$&$+2$  &  $+4$ &$+6$\\[0.5em]
$T_{2}^0$ &$\left(0~0~0~0~0~;\frac{-1}{3}~\frac{-1}{3}~\frac{-1}{3}
\right)(0~0~0~0~0~\frac{1}{2}~\frac{-1}{2}~0)'$ & L & $2_{\bar
1},2_{3}$ & $1,1$ & $S_{18}$&$-3$&$-4$& $-8$ & $-10$ &$-12$ \\[0.5em]
\hline &&&&& &&&&&\\[-1em]

$T_{1}^0$ &$\left(0~0~0~0~0~;\frac{-1}{6}~\frac{-1}{6}~\frac{-1}{6}
\right)(0~0~0~0~0~\frac{-3}{4}~\frac{1}{4}~\frac12)'$ & L &
$3_{3}$ & $1$ & $S_{19}$& $-1$&$0$&$+4$ &  $+6$ &$+8$\\[0.5em]
$T_{1}^0$ & $\left(0~0~0~0~0~;\frac{-1}{6}~\frac{-1}{6}~\frac{-1}{6}
\right)(0~0~0~0~0~\frac{1}{4}~\frac{-3}{4}~\frac12)'$ & L &$3_{3}$ & $1$
 & $S_{20}$ &$-1$&  $-2$&$-6$ & $-8$ &$-10$  \\[0.5em]
$T_{1}^0$ &$\left(0~0~0~0~0~;\frac{-1}{6}~\frac{-1}{6}~\frac{-1}{6}
\right)(0~0~0~0~0~\frac{1}{4}~\frac{1}{4}~\frac{-1}{2})'$ & L &
$\{1_{1},1_{3}\}$, $\{2_{3},1_{2}\}$,$6_{3}$ & $1,1,1$ & $S_{21}$& $-1$& $-1$&$-1$ &$-1$ &$-1$ \\[0.5em]
\hline &&&&& &&&&&\\[-1em]

$T_{7}^0$ &  $\left(0~0~0~0~0~;\frac{5}{6}~\frac{-1}{6}~\frac{-1}{6}\right)
 (0~0~0~0~0~\frac{-1}{4}~\frac{-1}{4}~\frac{1}{2})'$ & L & $2_{\bar 1}$ &
 $1$ & $S_{22}$ & $+1$& $+1$&$+1$ & $+1$ &$+1$\\[0.5em]
$T_{7}^0$ &  $\left(0~0~0~0~0~;\frac{-1}{6}~\frac{5}{6}~\frac{-1}{6}\right)
 (0~0~0~0~0~\frac{-1}{4}~\frac{-1}{4}~\frac{1}{2})'$ & L & $2_{\bar 1}$ &
 $1$ & $S_{23}$& $+1$& $+1$&$+1$ &  $+1$ &$+1$ \\[0.5em]
$T_{7}^0$ &   $\left(0~0~0~0~0~;\frac{-1}{6}~\frac{-1}{6}~\frac{5}{6}\right)
 (0~0~0~0~0~\frac{-1}{4}~\frac{-1}{4}~\frac{1}{2})'$ & L & $2_{\bar 1}$ &
 $1$ & $S_{24}$& $+1$&$+1$&$+1$ &$+1$ &$+1$ \\[0.5em]
\hline

\end{tabular}
\end{center}
\caption{All the SU(5)$\times$U(1)$_X\times$SU(5)$'\times$ SU(2)$'$ singlet states of Ref. \cite{HuhKimKyae09}.
  }\label{tab:singlets} }
\end{table}
\end{widetext}

Possible U(1) generators we illustrate, such that all Higgs singlets and doublets carry even quantum numbers and quark and leptons carry odd quantum numbers, are
\dis{
&Z_4 =  (0~0~0~0~0~ 2~ 2~ 0)\, (0~0~0~0~0~ -1~ 1~ 0)',\\
&Z_8 =  (0~0~0~0~0~ 2~ 2~ 0)\, (0~0~0~0~0~ -5~ 5~ 0)',\\
&Z_{10} =  (0~0~0~0~0~ 2~ 2~ 0)\, (0~0~0~0~0~ -7~ 7~ 0)',\\
&Z_{12} = (0~0~0~0~0~ 2~ 2~ 0)\, (0~0~0~0~0~ -9~ 9~ 0)',
}
which are mixtures of E$_8$ and E$_8'$ U(1)'s. These are the gauge U(1)'s whose normalization in E$_8$ or E$_8'$ part is such that
$Z^2_{i=4,8,10,12}(\rm E_8~or~E_8')=1$ \cite{ChoiKimBk}. All the integer entries imply appropriate directional cosines multiplied and
the charge operators $Z_i$ need not be normalized, \ie an appropriate factor is then multiplied such that the minimum nonzero magnitude of quantum numbers is 1. From Tables \ref{tab:T4Ord} and \ref{tab:singlets}, for R-parity the Higgs fields are chosen with even quantum numbers of $Z_4, Z_8, Z_{10}$, and $Z_{12}$. Thus, there results a matter parity $\Z_2$ and there exists a stable lightest $\Z_2$ odd particle in these theories.

\subsubsection{$\Z_{10}$}
To break the U(1) gauge group to $\Z_{10}$, refer to the $Z_{10}$ column of Table \ref{tab:singlets}.
In this specific GUT example, however, $\Z_{10}$ cannot be
obtained from the $Z_{10}$ column of Table \ref{tab:singlets} because 10 can be divided by 5(the center of SU(5)) and reduces to
$\Z_2$ \cite{LazShafi82}. Here, we just show the possibility of $Z_{10}$ quantum number, and if we really want to obtain $\Z_{10}$, we have to find out $Z_{50}$. In the following, we disregard the center of the GUT group.
If $s_9, s_{13}$, and $s_{18}$, carrying $\pm 10$ units of $Z_{10}$, develop VEVs but all the other singlets do not develop VEVs, then the vacuum discrete group is $\Z_{10}$. In addition, if the dimension-3 $W$ carries $Z_{10}=10$ and  $Z_{10}=0$ terms, the discrete group of dimension-3 $W$ is $\Z_{10}$. If higher order superpotential terms are not a multiple of 10, which will be the case generically, then $\Z_{10}$ is broken by these higher order terms. Below the electroweak scale the remaining discrete group cannot be larger than $\Z_2$ since the greatest common divisor of 2(the quantum number of $H_u$ and $H_d$) and 10(for $\Z_{10}$) is 2. Thus, there exists an unbroken $\Z_2$ and the lightest $\Z_2$ odd particle can be a CDM candidate.

\subsubsection{$\Z_{8R}$}
To break the U(1) gauge group to $\Z_{8}$, refer to the $Z_{8}$ column of Table \ref{tab:singlets}. If the dimension-3 $W$ carries $Z_{8}=10$ terms but no $Z_{8}=0$ term, then the discrete group of dimension-3 $W$ can be taken as $\Z_{8R}$.  To obtain $\Z_{8R}$,  $s_9, s_{13}$, and $s_{18}$ are required to obtain the GUT scale VEV, and all the other singlet VEVs are zero. Below the electroweak scale the remaining discrete group is $\Z_2$ since the greatest common divisor of 2(the quantum number of $H_u$ and $H_d$) and 8(for $\Z_{8R}$) is 2. Thus, there exists an unbroken $\Z_2$ and the lightest $\Z_2$ odd particle can be a CDM candidate.

\subsubsection{D- and F-flat directions}
In principle, the VEVs we have chosen must satisfy the D-flat and F-flat directions \cite{HuhKimKyae09, Lebedev08}. For example, let us consider the $\Z_{10}$ example of $\langle s_i \rangle\ne 0$ for $i=9,13, 10, 11, 12, 14, 15, 16$ and $\langle s_i \rangle= 0$  for $i\ne 9,13, 10, 11, 12, 14, 15, 16$. We have chosen the nonvanishing VEVs such that the D-flat condition can be satisfied.
To show the F-flat condition, we consider the following possible superpotential terms of the fields with nonvanishing VEVs,
\dis{
 -&M_1 s_9 s_{13} -\frac{M_2}{3}(s_{10} s_{14} + s_{11} s_{15}+s_{12} s_{16})\\
 & + \frac{1}{3M}(s_{10} s_{14} + s_{11} s_{15}+s_{12} s_{16}) s_9 s_{13},
}
where we simplified the parameters and also the VEVs  $ \langle s_i\rangle =V\, (i=9,13 )$ and $ \langle s_i\rangle =v\,(i=10, 11, 12, 14, 15, 16)$. The F-flat conditions lead to two independent equations,
\dis{
- \frac{M_2}{3} v +\frac{1}{3M}v V^2 =0,~~- M_1 V + \frac{1}{M}v^2 V=0 .
}
Thus, we can show the existence of a possible solution set, $v=\sqrt{ MM_1 }$ and $V= \sqrt{ M  M_2}$.
With $M_1\ll M_2$, we obtain $\Z_{10}$ between $\sqrt{ M  M_2}$ and $\sqrt{MM_1}$ and $\Z_2$ below $\sqrt{MM_1}$.

\subsection{Proton decay}

The R-parity forbids the dimension-3 superpotential term $u^cd^cd^c$ which  breaks the baryon number. But the dimension-4 superpotential terms, for example $qqq\ell$, are not forbidden. However, the dimension-4 superpotential term $\ell\ell H_uH_u$ is needed for neutrino masses via the seesaw mechanism. Forbidding $qqq\ell$ but allowing $\ell\ell H_uH_u$ have been studied under
`proton hexality' \cite{Dreiner06} and $\Z_{4R}$ \cite{LeeRatz11}. With the $Z_{4R}$ quantum numbers of Table \ref{tab:T4Ord}, the dimension-4 superpotential breaking $B$ is forbidden, but the dimension-4 superpotential terms for neutrino masses are allowed.
With $Z_{8R}$, the dimension-4 superpotential term breaking $B$ is forbidden, and also the dimension-4 superpotential terms for neutrino masses are not allowed. For neutrino masses, referring to the $Z_8$ column of Table \ref{tab:singlets}, one has to consider dimension-5 superpotential terms such as $(\{s_{14},\,s_{19}\}/M_P^2)\ell\ell H_uH_u$. With the GUT scale VEVs of these singlets, a correct order of neutrino masses can be obtained. In this case of $\Z_{8R}$ with GUT scale singlet VEVs, we have to check again the dimension-5 superpotential term breaking $B$. This can happen with $(s_{11}/M_P^2) qqq\ell$. So, requiring the VEVs of $s_{11}$ and $s_{14}$ sufficiently small, \ie $<10^{-7} M_P$, the effective $qqq\ell$ term is sufficiently suppressed.
In this sense, we have a hierarchy of VEVs, $s_i(i\ne 9, 11, 13, 14)\ll s_{11}\ll s_{14}\ll s_{9}, s_{13}$, satisfying the F- and D-flat conditions.
This scenario may be better than those completely forbidding proton decay, in the sense of observing proton decay in the near future.

\subsection{QCD axion}
For the approximate PQ symmetry in $\Z_{8R}$, one can consider many possibilities. But not writing down all the spectra here, it is difficult to present the scenario concretely. Here we present the idea only, with a hidden confining force for dynamical SUSY breaking. There exist heavy quarks. If the heavy quarks are assigned some PQ charges, then some singlets must carry the PQ charges since these singlet VEVs give masses to the heavy quarks. Let all the other fields carry zero PQ charges. Then, this PQ symmetry cannot be a gauge symmetry since the singlets must couple at some higher order to other fields such as to $H_u$ and $H_d$, allowed by gauge symmetry but forbidden by the PQ symmetry. So, the U(1)$_{\rm PQ}$ we introduced is an approximate one. If the string compactification does not introduce the anomalous U(1), we have (1) the model-independent axion \cite{MIaxion}\,\footnote{Here, we neglect the model-dependent axions which are probably very heavy \cite{WenWitten}, and the decay constant of the model independent axion is at the string scale \cite{ChoiKim84}.} and (2) the axion from the above PQ symmetry we introduced from the heavy quarks. This two axion system has the following feature.  The mass eigenstates are such that heavier axion (obtaining mass by the hidden sector gauge anomaly) has decay constant presumably at the intermediate scale $10^{10}-10^{12}\,\gev$ by the VEVs of singlets (the smaller VEV of the two scales) and the QCD axion has the decay constant raised to the string scale (the larger VEV of the two scales) \cite{TwoAxions}.
However, if the string compactification introduces the anomalous U(1) \cite{GreenSch84}, the anomalous U(1) becomes a global U(1) below the anomalous U(1) gauge boson mass. Again, the decay constant corresponding to the model-independent axion is expected at the GUT scale, and by a similar argument discussed above the QCD axion decay constant is expected at the GUT scale if a fine-tuning is not introduced. The basic reason for this difficulty results from another unbroken nonabelian gauge group at the intermediate scale, which can be circumvented in composite QCD axion scenarios \cite{KimNilles09}. It will be very useful if a fundamental QCD axion resolves this problem with discrete symmetries \cite{Kim14}.

\section{Conclusion}
We have presented a simple and explicit method of obtaining Abelian discrete groups $\Z_{N}$ from string compactification, by assigning VEVs to specified SM singlets. For the application, we need a full spectrum from string compactification. $\Z_N$ can be embedded in a gauged U(1) direction in the string compactification. In this case, we need not worry about discrete anomalies and obtaining the full spectrum is the requisite. To have a DM candidate a $\Z_2$ is required to be unbroken. If applied to global symmetries such as to U(1)$_{\rm PQ}$ and U(1)$_{R}$, the vacua are tilted. The discrete symmetries $\Z_{n+2}$ and $\Z_{nR}$
 from string compactification are also discussed.

\begin{acknowledgments}
I thank K.-S. Choi, I.-W. Kim and B. Kyae for useful discussions. This work is supported in part by the National Research Foundation (NRF) grant funded by the Korean Government (MEST) (No. 2005-0093841).
\end{acknowledgments}

\vskip 0.3cm



\begin{thebibliography}{99}

\def\plb#1#2#3{Phys.\ Lett.\ {\bf B#1} (#3) #2}
\def\prp#1#2#3{Phys.\ Rep.\ {\bf #1} (#3) #2}
\def\rmp#1#2#3{Rev. Mod. Phys.\ {\bf #1} (#3) #2}
\def\anrnp#1#2#3{Annu. Rev. Nucl. Part. Sci.\ {\bf #1} (#3) #2}
\def\npb#1#2#3{Nucl.\ Phys.\ {\bf B#1} (#3) #2}
\def\plb#1#2#3{Phys.\ Lett.\ {\bf B#1} (#3) #2}
\def\prd#1#2#3{Phys.\ Rev.\ {\bf D#1} (#3) #2}
\def\prl#1#2#3{Phys.\ Rev.\ Lett.\ {\bf #1} (#3) #2}
\def\jhep#1#2#3{JHEP\ {\bf #1} (#3) #2}
\def\jcap#1#2#3{JCAP\ {\bf #1} (#3) #2}
\def\zp#1#2#3{Z.\ Phys.\ {\bf #1} (#3) #2}
\def\epjc#1#2#3{Euro. Phys. J.\ {\bf C#1} (#3) #2}
\def\jpg#1#2#3{J. Phys.\ {\bf G#1} (#3) #2}
\def\ijmp#1#2#3{Int.\ J.\ Mod.\ Phys.\ {\bf #1} (#3) #2}
\def\mpl#1#2#3{Mod.\ Phys.\ Lett.\ {\bf A#1} (#3) #2}
\def\apj#1#2#3{Astrophys.\ J.\ {\bf #1} (#3) #2}
\def\nat#1#2#3{Nature\ {\bf #1} (#3) #2}
\def\sjnp#1#2#3{Sov.\ J.\ Nucl.\ Phys.\ {\bf #1} (#3) #2}
\def\apj#1#2#3{Astrophys.\ J.\ {\bf #1} (#3) #2}
\def\ijmp#1#2#3{Int.\ J.\ Mod.\ Phys.\ {\bf #1} (#3) #2}
\def\mpla#1#2#3{Mod.\ Phys.\ Lett.\ {\bf A#1} (#3) #2}
\def\nat#1#2#3{Nature\ {\bf #1} (#3) #2}
\def\npb#1#2#3{Nucl.\ Phys.\ {\bf B#1} (#3) #2}
\def\pthp#1#2#3{Prog.\ Theor.\ Phys.\ {\bf #1} (#3) #2}

\def\ibid#1#2#3{{\it ibid.} {\bf #1} (#3) #2}
\def\err#1#2#3{\ {\bf #1} (#3) #2\,(E)}

\bibitem{DiscreteR} S. Hamidi and C. Vafa,  \npb{279}{465}{1987};\\
 L. J. Dixon, D. Friedan, E. J. Martinec, and S. H.  Shenker,  \npb{282}{13}{1987}.

\bibitem{Kobayashi07} T. Kobayashi, H. P. Nilles, F. Fl\"oger, S. Raby, and M. Ratz, \npb{768}{135}{2007}
 [arXiv:hep-ph/0611020].

\bibitem{GravityGlobal}
S. M. Barr and D. Seckel,
     \prd{46}{539}{1992};\\
M. Kamionkowski and J. March-Russell,
          \plb{282}{137}{1992} [hep-th/9202003];\\
R. Holman, S. D. H. Hsu, T. W. Kephart, E. W. Kolb, R. Watkins, and L. M. Widrow,
     \plb{282}{132}{1992} [hep-ph/9203206];\\
B. A. Dobrescu,  \prd{55}{5826}{1997} [hep-ph/9609221].

\bibitem{Kim13} J. E. Kim, \prl{111}{031801}{2013} [arXiv:1303.1822 [hep-ph]].

\bibitem{PQ77} R. D. Peccei and H. R. Quinn,  \prl{38}{1440}{1977}; J. E. Kim, \prl{43}{103}{1979}; M. A. Shifman, V. I. Vainshtein, V. I. Zakharov, \npb{166}{493}{1980}; M. Dine, W. Fischler and M. Srednicki, \plb{104}{199}{1981}; A. P. Zhitnitsky, Sov. J. Nucl. Phys. {\bf 31}, 260 (1980).

\bibitem{KimRMP10}
For a recent review, see, J. E. Kim and G. Carosi, \rmp{82}{557}{2010} [arXiv: 0807.3125[hep-ph]].

\bibitem{StroWitten85}
A. Strominger and E. Witten, Commun. Math. Phys. {\bf 101} (1985) 341.

\bibitem{DiscrGauge89} L. M. Krauss and F. Wilczek,  \prl{62}{1221}{1989}.

\bibitem{Ibanez92} L. E. Ibanez and G. G. Ross, \plb{368}{3}{1992}.

\bibitem{BanksDine92} T. Banks and M. Dine, \prd{45}{1424}{1992}.

\bibitem{Preskill91} J. Preskill, S. P. Trivedi, F. Wilczek, and M. B. Wise, \npb{363}{207}{1991}.

\bibitem{Kim88} J. E. Kim, \plb{207}{434}{1988}; J. A. Casas, E. K. Katehou, and C. Munoz, \npb{317}{171}{1989}.

\bibitem{ChoiKW97} K. Choi, E. J. Chun and H. D. Kim, \prd{55}{7010}{1997} [arXiv:hep-ph/9610504].

\bibitem{LeeRatz11} H. M. Lee, S. Raby, M. Ratz, G. R. Ross, R. Schieren, K. Schmidt-Hoberg, and P. K. S. Vaudrevange,  \plb{694}{491}{2011} [arXiv:1009.0905]; \npb{850}{1}{2011} [arXiv:1102.3595 [hep-ph]]; M. Paraskevas and K. Tamvakis,  \prd{86}{015009}{2012} [arXiv:1205.1391 [hep-ph]].

\bibitem{Hall83} L. J. Hall and M. Suzuki, \npb{231}{419}{1983}; L. J. Hall, J. Lykken and S. Weinberg,
\prd{27}{2359}{1983}.

\bibitem{gminus2effSUSY} J. E. Kim, B. Kyae and H. M. Lee, \plb{520}{298}{2001} [arXiv:hep-ph/0103054].

\bibitem{IeffSUSY12} J. E. Kim, \prd{87}{015004}{2013} [arXiv:1208.5484[hep-ph]].

\bibitem{SSMkimjh07} J. E. Kim, J.-H. Kim and B. Kyae,  \jhep{0706}{034}{2007} [hep-ph/0702278].

\bibitem{Lebedev08}  O. Lebedev, H. P. Nilles, S. Raby, S. Ramos-Sanchez, M. Ratz, P. K. S. Vaudrevange, A. Wingerter, \prd{77}{046013}{2008} [arXiv:0708.2691 [hep-th]].

\bibitem{KamionPRP} For an LSP DM review, see, K. Griest and M. Kamionkowski, \prp{333}{167}{2000}.

\bibitem{AxinoDM}  For axino DM reviews, see, K.-Y. Choi, J. E. Kim and L. Roszkowski, [arXiv:1307.3330 [astro-ph.CO]];
 L. Covi and J. E. Kim, New J. of Phys. {\bf 11} (2009) 105003 
    [arXiv: 0902.0769[astro-ph/CO]].

\bibitem{HuhKimKyae09}  J.-H. Huh, J. E. Kim and B. Kyae, \prd{80}{115012}{2009} [arXiv: 0904.1108 [hep-ph]].

\bibitem{KimNilles84} J. E. Kim and H. P. Nilles, \plb{138}{150}{1984}.

\bibitem{ChoiKimIW07}  K.-S. Choi, I.-W. Kim and J. E. Kim, \jhep{0703}{116}{2007} [hep-ph/0612107].

\bibitem{ChoiKS09} K.-S. Choi, H. P. Nilles, S. Ramos-Sanchez, and P. K.S. Vaudrevange, \plb{675}{381}{2009}
[arXiv:0902.3070 [hep-th]].

\bibitem{KapplR09} R. Kappl, H. P. Nilles, S. Ramos-S\'anches, M. Ratz, K. Schmidt-Hoberg, and P. K. S.
Vaudrevange, \prl{102}{121602}{2009} [arXiv:0812.2120 [hep-th]].

\bibitem{KapplR10} F. Br\"ummer, R. Kappl, M. Ratz, and K. Schmidt-Hoberg, \jhep{04}{006}{2010} [arXiv:1003.0084 [hep-th]].

\bibitem{SMZ2Z2} M. Blaszczyk, S. Groot Nibbelink, M. Ratz, F. Ruehle, M. Trapletti, and P. K. S. Vaudrevange,
  \plb{683}{340}{2010} [arXiv:0911.4905 [hep-th]]

\bibitem{KimKyae07} J. E. Kim and B. Kyae, \npb{770}{47}{2007} [hep-th/0608086].

\bibitem{GreenSch84} M. B. Green and J. Schwarz, \plb{149}{117}{1984}.

\bibitem{PetersenB09} B. Petersen, M. Ratz, and R. Schieren, \jhep{0908}{111}{2009} [arXiv:0907.4049 [hep-ph]].

\bibitem{Munoz93} J. A. Casas and C. Munoz, \plb{306}{288}{1993}.

\bibitem{Dine09} M. Dine and J. Kehayias, \prd{82}{055014}{2009} [ arXiv:0909.1615 [hep-ph]]

\bibitem{NelsonSeib94} A. E. Nelson and N. Seiberg, \npb{416}{46}{1994} [arXiv:hep-ph/9309299].

\bibitem{ChoiKimBk} K.-S. Choi and J. E. Kim, \emph{Quarks and Leptons from Orbifolded Superstring} (Springer, Heidelberg, 2006).

\bibitem{IMNQ88} See also, L. E. Ibanez, J. Mas, H. P. Nilles, and F. Quevedo, \npb{301}{157}{1988}; T. Kobayashi, S. Raby, and R.-J. Zhang, \npb{704}{3}{2005} [hep-ph/0409098]; K.-S. Choi and T. Kobayashi, \npb{797}{295}{2008} [arXiv:0711.4894 [hep-th]].

\bibitem{LazShafi82} G. Lazarides and Q. Shafi, \plb{115}{21}{1982}; J. E. Kim, \prp{150}{1}{1987};
C. Csaki and H. Murayama, \npb{515}{114}{1998}.

\bibitem{Dreiner06} H. K. Dreiner, C. L\"uhn, and M. Thormeier, \prd{73}{075007}{2006} [arXiv:hep-ph/0512163]; S. F\"orste, H. P. Nilles, S. Ramos-Sanches, and P. K. S. Vaudrevange, \plb{693}{386}{2010} [arXiv:1007.3915 [hep-ph]]; K.-S. Choi and J. E. Kim, \prd{83}{065016}{2011} [arXiv: 1012.0847[hep-ph]].

\bibitem{MIaxion} E. Witten, \plb{149}{351}{1984}.

\bibitem{WenWitten} X. G. Wen and E. Witten, \plb{166}{397}{1986}.

\bibitem{ChoiKim84} K. Choi and J. E. Kim, \plb{154}{393}{1985}.

\bibitem{TwoAxions} J. E. Kim, \jhep{9905}{022}{1999} [arXiv:hep-th/9811509]; \jhep{0006}{016}{2000} [arXiv:hep-ph/9907528];
J. E. Kim and I.-W. Kim, \plb{639}{342}{2006} [arXiv:hep-th/0605256].

\bibitem{KimNilles09} J. E. Kim, \prd{31}{1733}{1985}; K. Choi and J. E. Kim, \prd{32}{1828}{1985};
 J. E. Kim and H. P. Nilles, \jcap{0905}{010}{2009} [arXiv: 0902.3610 [hep-th]].

\bibitem{Kim14} J. E. Kim, Work in progress.



\end{thebibliography}
\end{document}